# Fabrication and properties of MgB$_2$/AlO$_x$/MgB$_2$ tunnel junctions

K. Ueda, S. Saito, K. Semba and T. Makimoto

NTT Basic Research Laboratories, NTT Corporation,

3-1 Wakamiya, Morinosato, Atsugi-shi, Kanagawa 243-0198, Japan

M. Naito

Department of Applied Physics, Tokyo University of Agriculture and Technology,

2-24-16 Naka-cho, Koganei, Tokyo, 184-8588, Japan

Sandwich-type MgB$_2$ Josephson tunnel junctions (MgB$_2$/AlO$_x$/MgB$_2$) have been fabricated using as-grown MgB$_2$ films formed by molecular beam epitaxy. The junctions exhibited substantial supercurrent and a well-defined superconducting gap ($\Delta$=2.2~2.3mV). The superconducting gap voltage ($\Delta$) agrees well with those of the smaller gap in the multi-gap scenario. The I$_c$R$_N$ product is 0.4-1.3 mV at 4.2K, but approaches to 2.0 mV at 50 mK. The I$_c$ has peculiar temperature dependence far from the Ambegaokar-Baratoff formula. It rapidly decreases with temperature, and disappears above T = 20 K, which is much lower than the gap closing temperature. Interface microstructure between AlO$_x$ and MgB$_2$ were investigated using cross-sectional transmission electron microscopy to clarify the problems in our tunnel junctions. There are poor-crystalline MgB$_2$ layers and/or amorphous Mg-B composite layers of a few nanometers between AlO$_x$ barrier and the upper MgB$_2$ layer. The poor-crystalline upper Mg-B layers seem to behave as normal metal or deteriorated superconducting layers, which may be the principal reason for all non-idealities of our MgB$_2$/AlO$_x$/MgB$_2$ junctions.

1. Introduction

Recently discovered superconducting magnesium diboride, $MgB_2$, [1] is an attractive alternative for superconducting electronics applications. Its high transition temperature ($\sim 40$ K) enables the use of economical compact cryocoolers to maintain an operation temperature 20-25 K, which is a great advantage for Nb-based superconducting electronics that required the liquid He or large-scale cryocoolers. Josephson junctions are the key elements of superconducting electronic circuits due to their phase-sensitive supercurrent and highly nonlinear current-voltage characteristics. Using the fabrication technology established in the early 1980's, Nb-based Josephson junctions can be uniformly and reproducibly fabricated using thin $Al_2O_3$ layer as an insulating barrier. At present, Nb-based microprocessors using single-flux quantum (SFQ) logic have proven to operate successfully over 20 GHz. These prototype superconducting microprocessors, integrated as many as $\sim 10,000$ Josephson junctions. In contrast, we have not yet seen any great promise for electronic circuits using copper oxide high-$T_c$ superconductors (HTS), which are supposed to operate above the liquid nitrogen temperature. This is because, after almost 20 years since their discovery, no technology for fabricating HTS Josephson junctions uniformly has yet been established. Therefore, the expectations are large for $MgB_2$ in spite of it not having a $T_c$ as high as HTS. In this article, we report the fabrication and properties of $MgB_2$-based Josephson junctions. Based on the results, the potential of $MgB_2$ for future superconducting electronics applications is discussed.

$MgB_2$ is a binary line compound with a simple crystal structure (the so-called $AlB_2$ structure) and has the highest $T_c$ ($\sim 40$ K) among intermetallic compounds [1]. $MgB_2$ has fewer material complexities, less anisotropy, and a longer coherence length ($\xi_{ab}= 10 \pm 2$ nm, $\xi_c= 2$-3 nm [2]) than high-$T_c$ cuprates, which are composed of more than three cations plus oxygen and have complex crystal structures. The material simplicity of $MgB_2$ may be its largest advantage for establishing a robust technology to fabricate high-quality Josephson junctions. Various types of superconducting $MgB_2$ junctions have already been reported [3-8]. Initial attempts using point contact junctions have already shown promising results, such as clear quasi-particle spectra with a well-defined superconducting gap of ~2.0 mV, excellent SQUID performance with flux noise and field noise as low as 4 $\mu\Phi_0 Hz^{-1/2}$, where $\Phi_0$ is the flux quantum, and 35 $fTHz^{-1/2}$ [3]. Furthermore, it has been shown that nanobridges patterned by a focused-ion beam (FIB) down to < 100 nm have outstanding critical current density approaching to $10^7$ A/cm$^2$ [6], indicating that the superconductivity of $MgB_2$ survives after nanostructure formation even at a length scale of < 100 nm. After these initial successes, the fabrication of sandwich-type tunnel junctions was attempted, since they are very important from the viewpoint of the reproducible integration of superconducting circuits. In early attempts, tunnel junctions comprising one $MgB_2$ electrode and one low-$T_c$ superconductor (LTS, such as Nb, NbN, V) or normal metal (such as Au, Ag) were fabricated [9-11]. The characteristics of these junctions were rather close to those of typical superconductor(1)/insulator/superconductor(2) (SIS') or superconductor/insulator/normal-metal (SIN) junctions with a well-defined superconducting gap and small subgap conductance. An almost ideal Fraunhofer pattern was also observed for NbN/AlN/$MgB_2$ junctions [10]. These early results have indicated that $MgB_2$ junctions more closely resemble LTS junctions than HTS junctions, which very rarely show clear tunneling characteristics, except for intrinsic ones.

Very recently Shimakage et al. have reported the first fabrication of all $MgB_2$ SIS junctions using AlN insulating barrier [12]. In their junction fabrication, both $MgB_2$ and AlN layers were formed by sputtering. Subsequently, we have shown that high-quality $MgB_2$-SIS junctions can be fabricated using an $AlO_x$ barrier [13]. In our junction fabrication, we formed both $MgB_2$ and $AlO_x$ *all-in-situ* by molecular beam epitaxy, which is the key to obtaining high-quality SIS junctions. Our junctions exhibited substantial supercurrent ($I_cR_N$ products of over 1 mV at the lowest temperature measured), a well-defined superconducting gap ($\Delta$=2.2~2.3mV), and clear Fraunhofer patterns. Here, we present more detailed descriptions of our fabrication of $MgB_2/AlO_x/MgB_2$ junctions, and the junction characteristics. We also present a microscopic structural characterization of the interfaces between the $MgB_2$ and the $AlO_x$ barrier layers by high-resolution transmission electron microscopy.

2. Experimental

$MgB_2$ films for tunnel junctions were prepared by the molecular beam epitaxy (MBE) method. The details of the $MgB_2$ film preparation are described [14, 15]. Briefly, we deposited Mg and B metals on sapphire-C substrates heated to 240–270°C by electron guns in an ultra-high vacuum chamber (basal pressure < $10^{-9}$ Torr). The evaporation rate of each element was controlled by electron impact emission spectrometry (EIES) via feedback loops to the electron guns. The flux ratio of Mg to B was set to a slightly larger value than 1:2 (typically 1.3:2) to compensate for the loss of Mg due to reevaporation from the film surface. The growth rate was 1.5 – 2 A/sec, and the thickness of each $MgB_2$ layer was 100-150 nm. The resultant $MgB_2$ films were c-axis oriented and had a superconducting transition temperature of 32 to 35 K.

Four layers, $MgB_2/AlO_x/MgB_2$ and the Au protective layer, were sequentially deposited *in-situ*. The $AlO_x$ barrier was formed by depositing a few nanometers of Al metal and oxidizing it in a load-lock chamber for 30 min in 100 Torr of pure oxygen. The deposition rate of Al layer was 0.1-0.2 nm/s, and the Al layer thickness ($d$) was varied from 0.4 to 5 nm. The deposition temperature ($T_s$) for the Al overlayer is one of the important parameters for obtaining good junctions. For $T_s$ higher than 200°C, we obtained shorted or leaky junctions with high probability. This is due to poor Al coverage on a lower $MgB_2$ electrode, presumably owing to large grain size. In general, the lower the $T_s$ is, the higher the yield to obtain good junction characteristics. So far, we have deposited Al at ambient temperatures (below $T_s \sim 80°C$), although we believe that intentional cooling of substrate may improve the yield further.

Junctions of 25-, 50- and 100-μm square in area were delineated by standard photolithography and Ar ion milling. A schematic diagram of our fabrication process is shown in Fig. 1. The current- voltage ($I$-$V$) characteristics were measured using the four-probe method in a standard liquid-He cryostat or sometimes in a dilution refrigerator. The Josephson current ($I_c$) so far observed was as small as a few to a few ten microamperes. In order to measure such low $I_c$, special attention was paid to reducing circuit noise using RCR (R: resistance, C: capacitance) filter at the sample temperature and low-pass π-type filters (cutoff frequency: 1.9 MHz) at room temperature.

We have also investigated the microstructure near the interface region of our $MgB_2/AlO_x/MgB_2$ junctions by high-resolution transmission electron microscopy (TEM) and electron micrograph diffraction patterns.

## 3. Results and Discussion

Figure 2 shows current-voltage (I-V) characteristics and differential conductance curves of $MgB_2/AlO_x/MgB_2$ junctions as a function of bias voltage (dI/dV-V) measured at 4.2 K. The dI/dV-V curves were digitally calculated from the I-V characteristics. The thickness of the Al overlayers was 1.4 nm for both junctions, and the junction areas were 100-μm square for junction A [Fig. 2(a)] and 25-μm square for junction B [Fig. 2(b)]. These junctions showed supercurrent and clear quasiparticle tunneling characteristics. The superconducting gap at $2\Delta_S$ was clearly observed at around 4.5 mV, which agrees well with the smaller gap ($\Delta_S$) in the multi-gap scenario for $MgB_2$ [16, 17]. The subgap leakage current is also small, and almost negligible below 2.0 mV ($\sim\Delta_S$), especially in junction B. The $R_{sg}/R_N$ (subgap resistance at 2.0 mV divided by normal resistance) measured at 4.2 K is 5.4 for junction A and 21.8 for junction B. The superconducting current ($I_c$) is observed in both junctions. The $I_c$ and normal resistance ($R_N$) at 4.2 K are 12.5 μA and 61 Ω for junction A and 0.4 μA and 830 Ω for junction B.    The $I_cR_N$ products at 4.2 K are ~0.8 mV for junction A and ~0.4 mV for junction B.    The $J_c$ (Josephson current density: $I_c$ divided by junction area) is estimated to be 0.125 A/cm$^2$ for junction A and 0.064 A/cm$^2$ for junction B at 4.2 K, which is quite small.    The $I_c$ becomes substantially larger at the lowest temperature (~50 mK) measured.

Figure 3 shows the dc magnetic field dependence of $I_c$ in junction A measured at 45 mK.    The magnetic field was applied parallel to the surface of the junction.    The $I_c$ was suppressed as the external magnetic field increased and showed a familiar Fraunhofer pattern, which agrees well with the calculated curve (solid line).    This observation indicates that the supercurrent density ($J_c$) is fairly uniform over the junction area.

Figure 4 shows the effect of the Al layer thickness (*d*) on the junction characteristics of MgB$_2$/AlO$_x$/MgB$_2$, where *d* is (a) 0.4 nm for junction C and (b) 5.0 nm for junction D, which should be compared with junction A and B with *d* = 1.4 nm.    Junction C exhibits SIS tunneling characteristics with $I_c$ (40.8 μA at 4.2 K) that are two orders of magnitude larger than those of junction B with the same area.    The $I_cR_N$ product at 4.2 K and $J_c$ for junction C is ~1.3 mV and 6.5 A/cm$^2$.    In spite of larger $I_cR_N$ product and $J_c$, the superconducting gap voltage of junction C is slightly smaller (4.0 mV), and there is also extra conductance below the superconducting gap voltage, indicating that electron transport other than tunneling is contributing.    However, it is rather surprising that MgB$_2$/AlO$_x$/MgB$_2$ junctions still exhibit SIS characteristics even with an Al overlayer as thin as 0.4 nm.    It is unlikely that such a thin Al overlayer fully covers the MgB$_2$ surface.    We speculate that pinholes in AlO$_x$ barrier layers may be filled up by the so-called "native" barrier (presumably MgO), which is produced in the oxidation process of Al overlayers.    In contrast, junction D with d=5.0 nm exhibits neither SIS characteristics nor Josephson current.    The dI/dV characteristics of this junction are rather SIN-like with a threshold voltage of 3.0 mV.    The gap value is between $\Delta_S$ and $2\Delta_S$, indicating that there may be a deteriorated layer either between the lower $MgB_2$ and barrier or between the barrier and upper $MgB_2$.    We have demonstrated before that $Au/Al_2O_3/MgB_2$ junctions exhibit typical SIN characteristics with a superconducting gap of 2.0-2.5 mV [18].    Based on this result, we suspect that a deteriorated superconducting layer or even a normal layer exists at the interface between the barrier layer and upper $MgB_2$ in junction D.    This would mean that the range for the Al overlayer thickness to obtain good $MgB_2/AlO_x/MgB_2$ junctions is narrow.    Compared with $Nb/AlO_x$-$Al/Nb$ junctions, which allow a rather wide range of Al thickness from 1 to 10 nm.

The junction parameters ($I_cR_N$, $J_c$, $R_NA$, etc.) of $MgB_2/AlO_x/MgB_2$ junctions are summarized in Table 1.    The $I_cR_N$ product of the junctions is the same order as the superconducting gap, which is in contrast to the situation in HTS junctions.    The $I_cR_N$ product of $YBa_2Cu_3O_{7-\delta}$ is at most a few mV, which is one order of magnitude lower than the superconducting gap (~30 mV).    As will be explained below, the disordered interface between the barrier and upper $MgB_2$ may be the primary reason that the $I_cR_N$ is somewhat smaller than the calculated value (3-4 mV, based on the smaller one of the two gaps) [19].    In spite of the substantial $I_cR_N$ product, the $J_c$ values of our $MgB_2/AlO_x/MgB_2$ junctions are quite small.    The small $J_c$ may come from our Al layer oxidation condition.    In $Nb/AlO_x$-$Al/Nb$ Josephson junctions, the $J_c$ becomes as high as $10^4$ A/cm$^2$ by controlling the oxidation pressure of Al [20, 21].    Typical oxidation pressure ($P_{O2}$) for Al oxidation in Nb-SIS junctions is 0.1-1Torr for 10-30 min, which is much lower than our typical oxidation condition ($P_{O2}$=100 Torr for 30 min).    We tried fabricating $MgB_2$-SIS junctions in almost the same oxidation condition as that used for Nb-SIS junction ($P_{O2}$=1 Torr for 10 min); however, the $MgB_2$ junctions become leaky and showed SNS-like behavior.    It is important to find the optimal $AlO_x$ barrier layer fabrication conditions for $MgB_2$-SIS junctions in order to obtain junctions with higher $J_c$.    The $R_NA$ (resistance for a junction with area A=1cm$^2$) values of junctions A and B, which have the same $AlO_x$ layer thickness, are 6.1$\times$ 10$^{-3}$ $\Omega$/cm$^2$ and 5.2$\times$ 10$^{-3}$ $\Omega$/cm$^2$, respectively.    The values agree well with each other, indicating fair reproducibility.

The temperature dependence of $I_c$ ($I_c$-T) for a junctions A and C is plotted in Fig. 5(a). The $I_c$ remained finite only up to around 20 K. It should be compared with the temperature dependence of the superconducting gap voltage ($\Delta$). The temperature dependence of $\Delta$ is shown in Fig. 5(b). The gap remained finite up to ~ 30 K, which is close to the resistive $T_c$ of our $MgB_2$ films [14, 15]. The $I_c$ also shows peculiar temperature dependence: it rapidly decreases (exponential-like) with increasing temperature, which is far from the Ambegaokar-Baratoff formula predicted for Josephson tunnel junctions. Such behavior is often observed in junctions with a superconductor-normal metal (SN) boundary [22]. Hence, this dependence indicates the existence of a normal-metal layer between superconducting $MgB_2$ layers and barriers in the junctions.

Next, we turn to our TEM investigation of $MgB_2/AlO_x/MgB_2$ junctions. Figure 6 is a cross-sectional TEM image demonstrating the initial growth of $MgB_2$ layers on a sapphire-C substrate. The interface between sapphire-C and $MgB_2$ layer is clean and sharp with neither an intermediate layer nor interdiffusion between them, indicating good initial growth of $MgB_2$ layer. These TEM results should be compared with those for $MgB_2$ films formed on sapphire-C by hybrid physical-chemical vapor deposition (HPCVD) at much higher temperatures (above 700°C), which exhibit MgO intermediate layers arising from the reaction between Mg and $Al_2O_3$ [23]

Figure 7 shows a cross-sectional TEM image of the interface region of an $MgB_2/AlO_x/MgB_2$ junction with Al-overlayer thickness of 1.4 nm. The wettability of Al to the $MgB_2$ seems to be very good, and the interface between lower $MgB_2$ layers and $AlO_x$ is quite sharp. However, there might be a very thin native-oxide layer (< ~0.5 nm) between the $AlO_x$ and lower $MgB_2$ layer. In contrast, the initial growth of upper $MgB_2$ layers is not good. There are poor-crystalline $MgB_2$ layers and/or amorphous Mg-B composite layers of a few nanometers between $AlO_x$ and upper $MgB_2$ layer. In order to take a closer look at the crystallinity of the lower and upper interface region of $MgB_2$, selected-area electron diffraction (SAED) patterns were obtained (Fig. 8) with the incident electron beam along the [10$\underline{1}$0] direction of an $Al_2O_3$ substrate.    The diffraction spots for the lower interface region are sharp [Fig. 8(a)], whereas those for the upper interface region are somewhat streaky [Fig.8(b)], which agrees with the TEM image in Fig. 7.    The diffraction pattern confirms that $MgB_2$ films are c-axis oriented with $(0001)_{MgB2}$ // $(0001)_{Al2O3}$.    With regard to the in-plane epitaxial relationship, we did see double domains with dominant [11$\underline{2}$0]$_{MgB2}$ // [10$\underline{1}$0]$_{Al2O3}$ and subsidiary [10$\underline{1}$0]$_{MgB2}$ // [10$\underline{1}$0]$_{Al2O3}$ in both the lower and upper $MgB_2$ layers.    The dominant grains grow in such a way that the hexagonal $MgB_2$ lattice is rotated by 30° to match the hexagonal lattice of sapphire.

The poor-crystalline $MgB_2$ layers and/or amorphous Mg-B composite layers between $AlO_x$ and upper $MgB_2$ seem to behave as a normal-metal or reduced-$T_c$ superconducting layer and severely affect the junction characteristics.    This may be a principal reason for all non-idealities of our $MgB_2/AlO_x/MgB_2$ junctions.    We should mention that a similar problem has been pointed out in $Nb/AlO_x$-Al/Nb junctions.    Poor-crystalline upper Nb has a lower $T_c$ and reduced superconducting gap, which contributes to finite subgap conductance above single $\Delta$ and broadens the threshold voltage (~2$\Delta$) for quasi-particle current.    Good initial growth of upper $MgB_2$ layers may be the key to improving $MgB_2/AlO_x/MgB_2$ junctions further.    At the same time, it is important to find a better barrier material than $AlO_x$.    It has been established that aluminum oxide is the best barrier material for Nb junctions in that Al not only wets Nb well but also has stronger oxygen affinity than Nb, which prevents Nb oxidation.    In the case of $MgB_2$, Mg has stronger oxygen affinity than Al, and thereby oxygen in $AlO_x$ barrier may be captured by Mg during upper $MgB_2$ growth.    In principle, if one uses elemental oxides as a barrier for $MgB_2$ junctions, the elements have to have stronger oxygen affinity than Mg.    The principle is identical to the one for choosing gate-insulator oxides for silicon.    Alternatively, we should also take a look at non-oxide barriers like AlN, which was demonstrated before to work well by Shimakage et al [12].

## 4. Summary and prospects

In summary, $MgB_2/AlO_x/MgB_2$ Josephson tunnel junctions were fabricated using MBE-grown $MgB_2$ films.   The keys to obtaining good junctions are (1) low deposition temperature for the Al layer for high coverage, and (2) appropriate Al layer thickness (~1 nm).   The junctions showed typical SIS characteristics with substantial supercurrent and a well-defined superconducting gap ($\Delta$= ~2.3 mV).   The superconducting gap voltage agrees well with that of the smaller gap ($\Delta_s$).   The $I_cR_N$ product of the junctions is 0.4 to 1.3 mV at 4.2 K and approached 2 mV at 50 mK.   The junctions so far obtained have high $R_N$ and small $I_c$, which is partly because the oxidation condition of the Al layer is not optimized.   The interface microstructure between the $AlO_x$ and $MgB_2$ films was investigated using cross-sectional transmission electron microscopy.   There are poor-crystalline $MgB_2$ layers and/or amorphous Mg-B composite layers of a few nanometers between the $AlO_x$ barrier and upper $MgB_2$ layer.   The poor-crystalline upper Mg-B layers seems to behave as normal metal or deteriorated superconducting layers, which may be the principal reason of all non-idealities of our $MgB_2/AlO_x/MgB_2$ junctions.   A key to improving $MgB_2/AlO_x/MgB_2$ junctions is to achieve better initial growth of the upper $MgB_2$ layers.

In terms of superconducting electronics applications, high-quality as-grown films and Josephson junctions are two prerequisites.   In the case of high-$T_c$ cuprates, the technology for growing high-quality epitaxial thin films was established in the first 2~3 years, but no reliable fabrication technology for high-quality multilayer Josephson junctions has been established after almost the 20 years since the discovery.   Electronics applications with high-$T_c$ cuprates have therefore been limited to devices using plain films, such as passive RF devices and SQUIDs with grain boundary Josephson junctions.   With regard to the tremendous difficulty in fabricating

Josephson junctions with high-$T_c$ cuprates, we pointed out an intrinsic and serious problem of cuprates: the redox reaction at the interface between cuprates and other materials.   Oxygen in cuprates is loosely bound and easily extracted by contact with other materials.    This leads to serious degradation of superconductivity at the interface of cuprates.    Without precise control of interface oxygen, it is very hard to obtain reproducible results in any type of high-$T_c$ junctions (not only SIS but also SIN or SIS').

In contrast, for $MgB_2$, we showed here that the two prerequisites are now essentially at hand. Hence, we believe that $MgB_2$ may be promising for superconducting electronics applications in spite that $T_c$ is not as high as that of high-$T_c$ cuprates.    One problem that should be mentioned is that $MgB_2$ is a two-gap superconductor with the larger gap $\Delta_L$ of ~6 mV ($2\Delta_L/k_B T_c \sim 4$) and the smaller gap $\Delta_S$ of ~2 mV ($2\Delta_S/k_B T_c \sim 1.2$).    This feature is interesting as science, but may become an obstacle for technology.    For high-frequency superconducting devices such as SIS mixers, the upper limit ($f_g$) of the response frequency is proportional to the superconducting energy gap.    The $f_g$ for existing devices with Nb ($\Delta = 1.5$ mV) is ~700 GHz.    In order to achieve the maximum performance from $MgB_2$ devices, we needs to extract the larger gap.    Then, $f_g$ will be ~3 THz.    As seen in this article, however, our $MgB_2$ junctions exhibited only the smaller gap (2~2.5 mV); they showed no signature indicating the larger gap.    The boron $sp^2$ ($\sigma$) band with 2D Fermi surface is responsible for the larger superconducting gap.    Extracting it may require the fabrication of Josephson junctions with a geometry that puts the supercurrent in the ab-plane as in high-$T_c$ ramp edge junctions.    An alternate way, as predicted by theories, would be to suppress the gap anisotropy by interband scattering via artificial introduction of impurities without reducing $T_c$ much.

Acknowledgement

    The authors thank Drs. H. Sato, H. Yamamoto, S. Karimoto and H. Shibata for fruitful discussions,

Drs. S. Mizuno, T. Iizuka and T. Mitate (NTT-AT) for TEM observations, and Drs. K. Torimitsu and H. Takayanagi for their support and encouragement throughout the course of this study.

Figure captions

Fig. 1: Schematic diagram of the fabrication process for $MgB_2/AlO_x/MgB_2$ junctions. Small drawings in (b), (c) and (e) are top views of the junctions.

Fig. 2: Current-voltage characteristics and differential conductance curves as a function of voltage for $MgB_2/AlO_x/MgB_2$ junctions with an area and Al overlayer of (a) 1.4 nm and 100-μm square (junction A) and (b) 1.4 nm and 25-μm square (junction B) measured at 4.2 K.

Fig. 3: Experimental and calculated dc magnetic field dependence of the superconducting critical current ($I_c$) of junction A measured at 45 mK. Here, $\Phi_0$ is a flux quantum, $\Phi_0 = hc/2e$.

Fig. 4: Current-voltage characteristics and differential conductance curves as a function of voltage for $MgB_2/AlO_x/MgB_2$ junctions with an area and Al overlayer of (a) 0.4 nm and 25-μm square (junction C) and (b) 5.0 nm and 100-μm square (junction D) measured at 4.2 K.

Fig. 5: (a) Temperature dependence of maximum $I_c$ of $MgB_2/AlO_x/MgB_2$ junctions with an area of 25-μm square, and an Al overlayer of 1.4 nm (　; junction A) and 0.4 nm (　; junction C).

(b) Temperature dependence of superconducting gap voltage ($\Delta$) for junction A (Al-overlayer: 1.4nm. Junction area 100-μm square). The solid line shows the calculated temperature dependence of $\Delta$.

Fig. 6: Cross-sectional TEM image of interfaces between the bottom $MgB_2$ layer and the sapphire -C substrate.

Fig. 7: Cross-sectional TEM image of $MgB_2/AlO_x/MgB_2$ interfaces fabricated on sapphire -C substrates.

Fig. 8: Magnified electron diffraction patterns of the (a) bottom and (b) upper $MgB_2$ layer of the $MgB_2/AlO_x/MgB_2$ junction.

-----------------

Table 1: Junction parameters of $MgB_2/AlO_x/MgB_2$ tunnel junctions shown in Fig. 1.

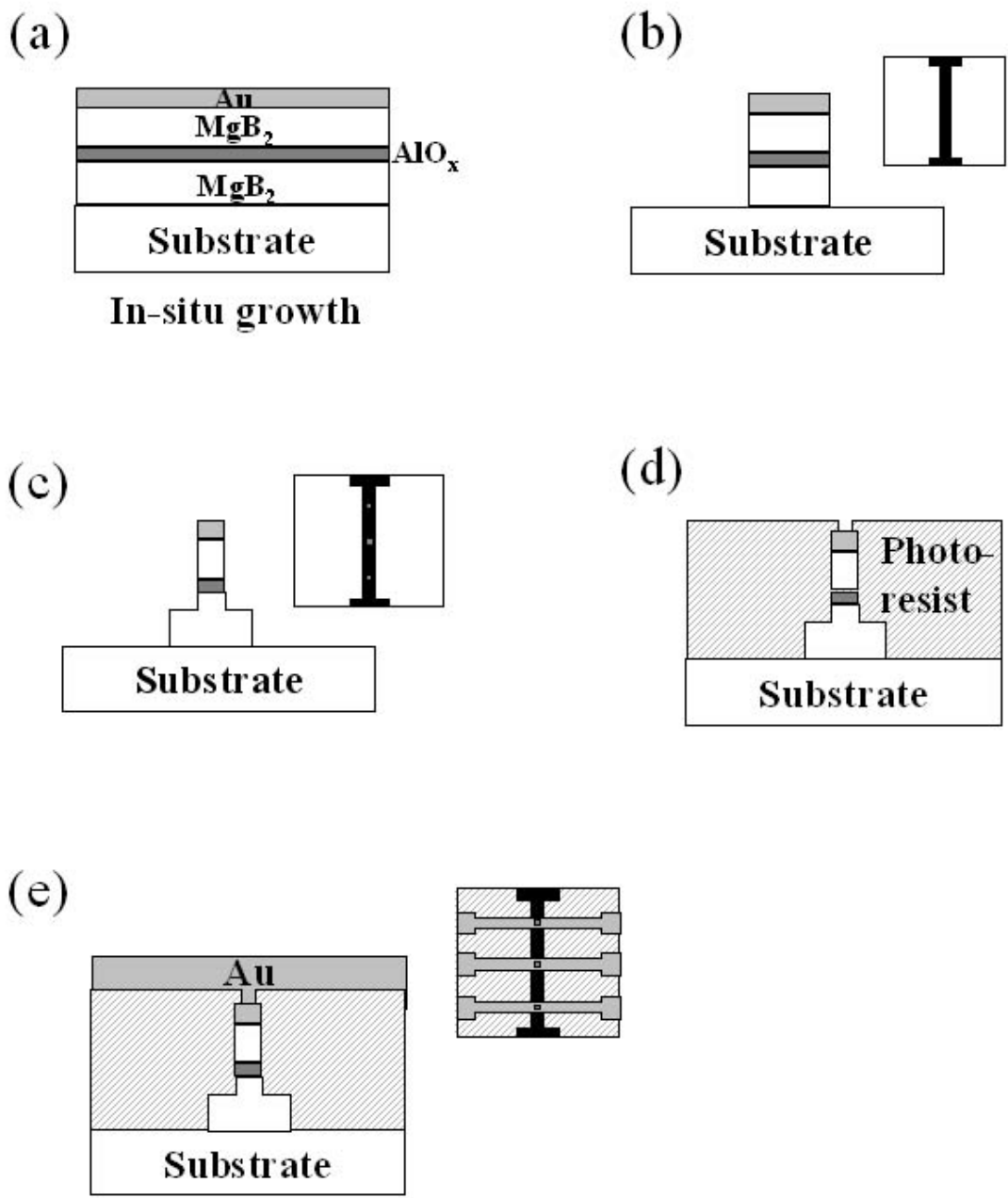

Fig. 1

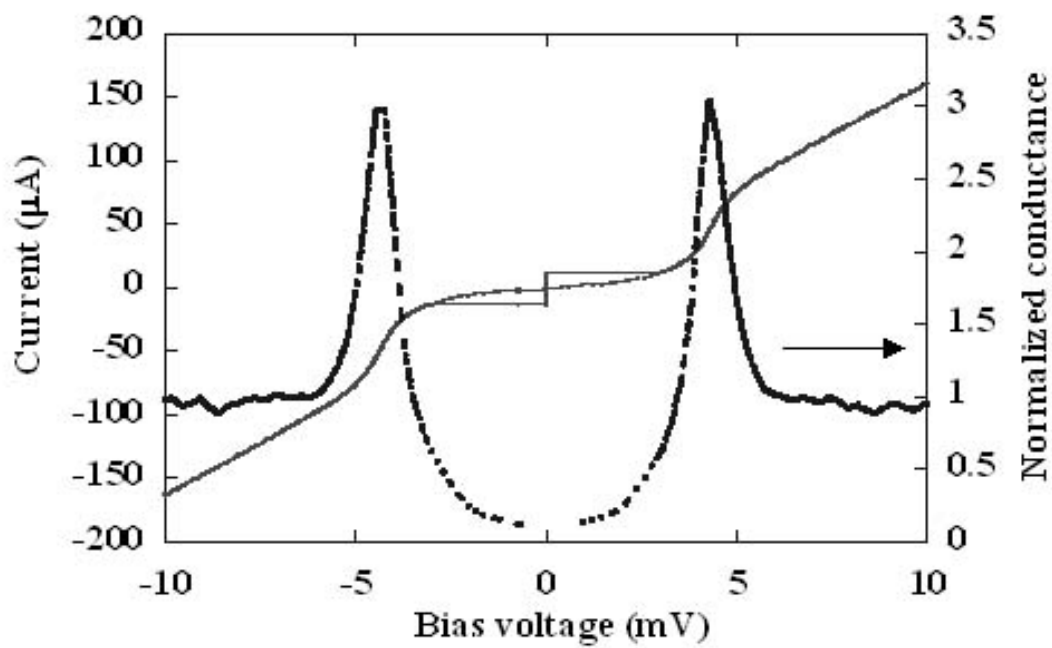

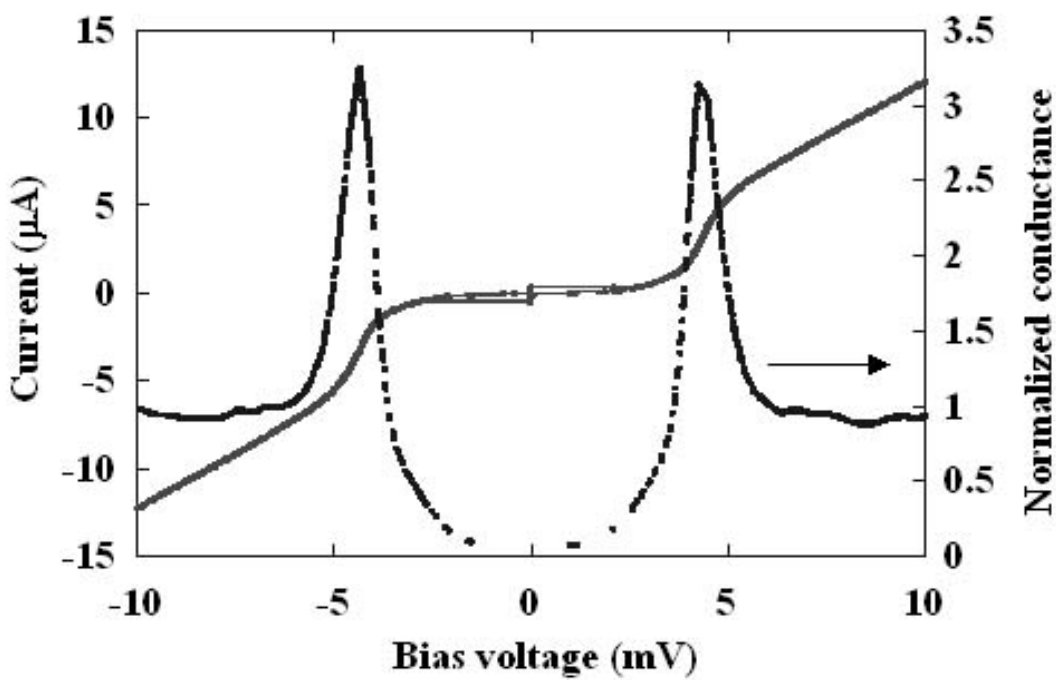

Fig. 2

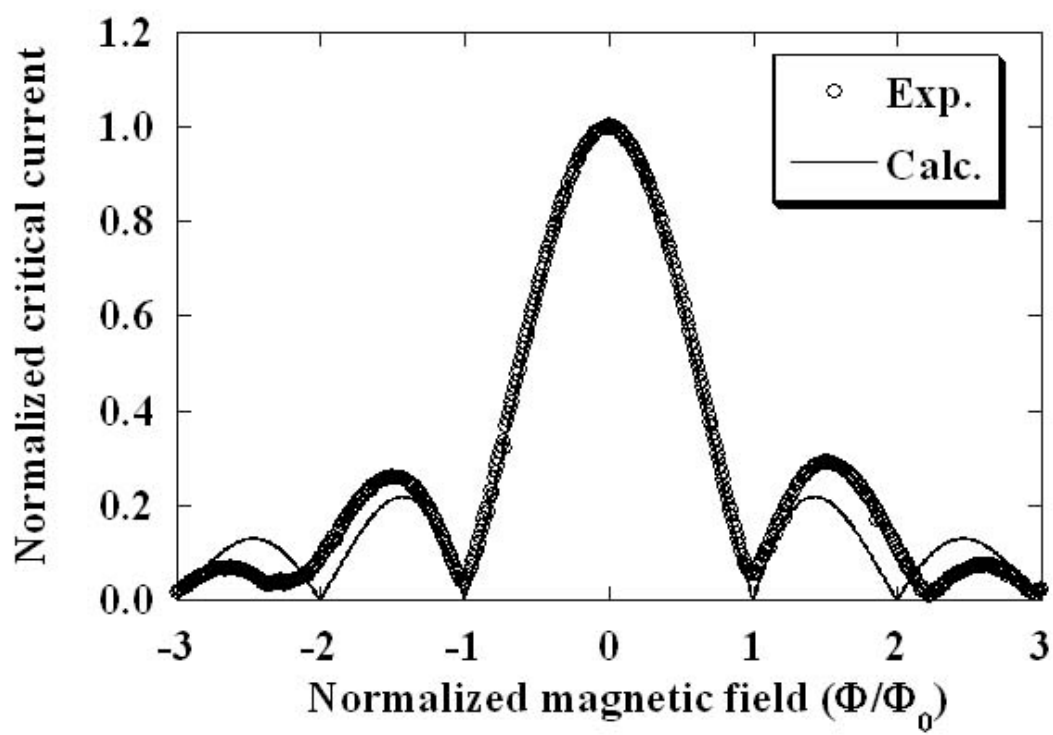

Fig. 3

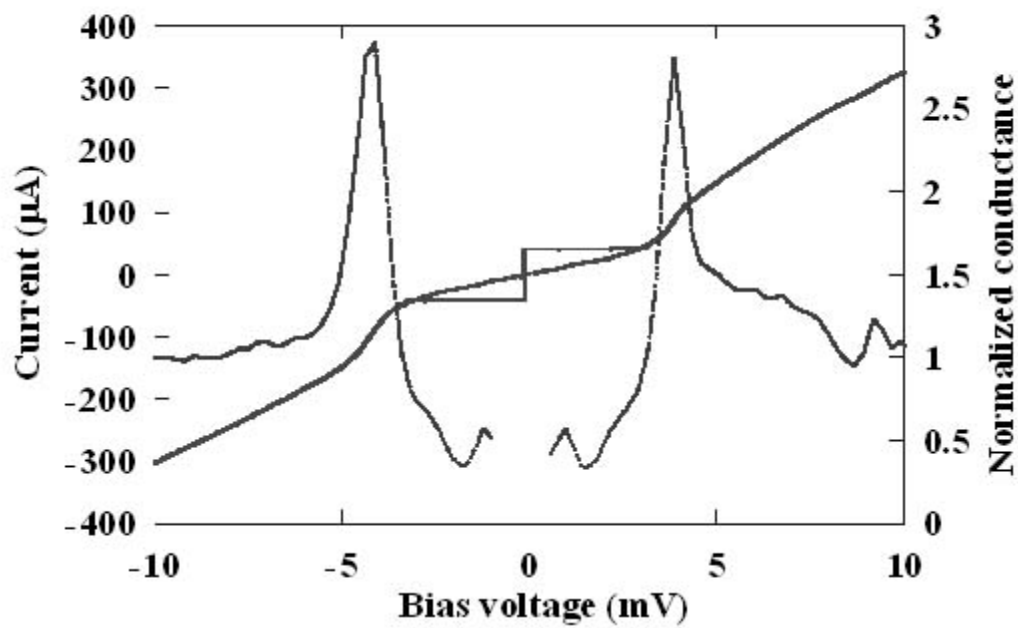

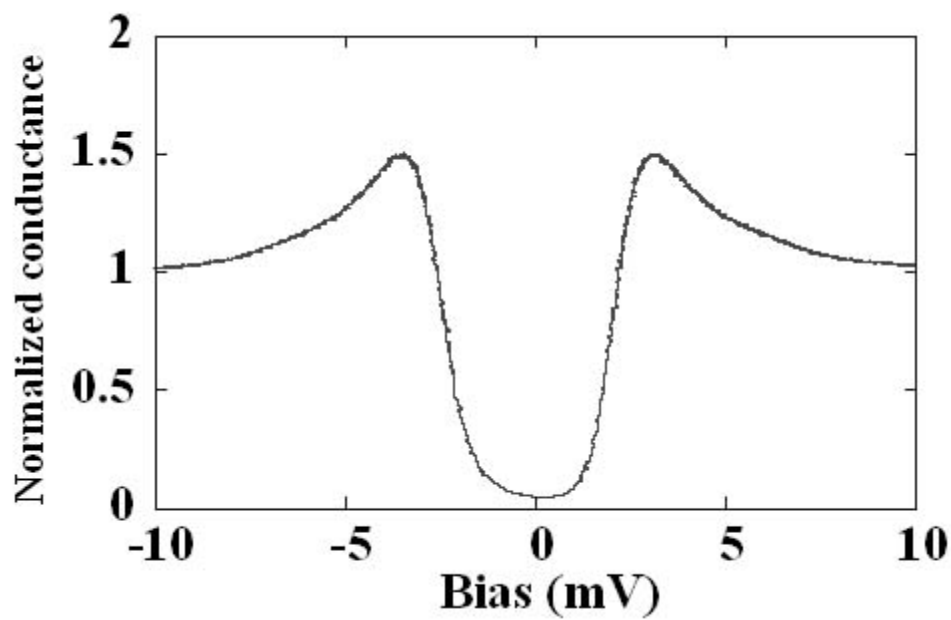

Fig. 4

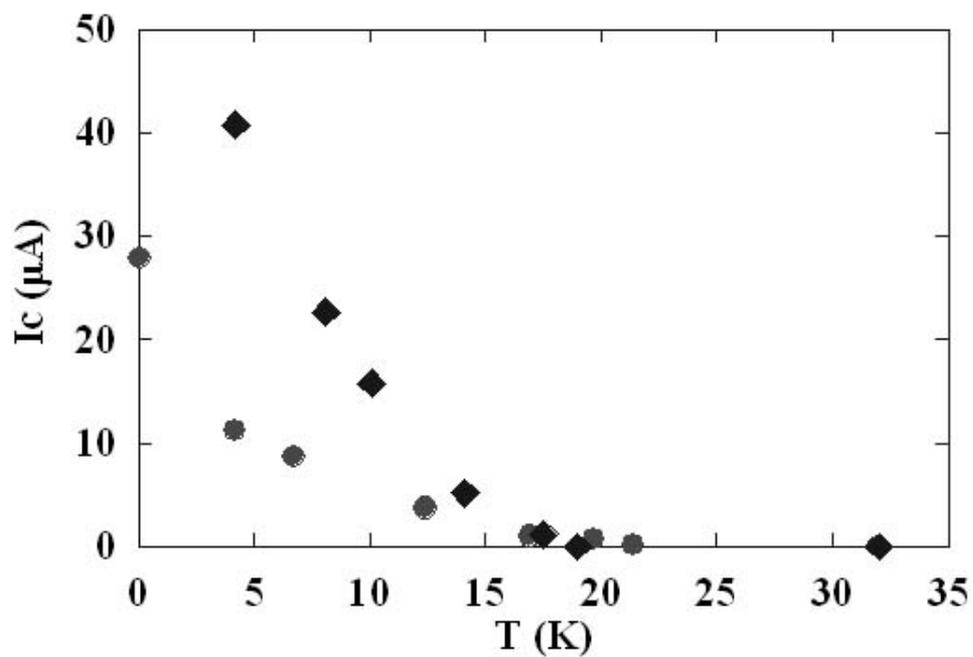

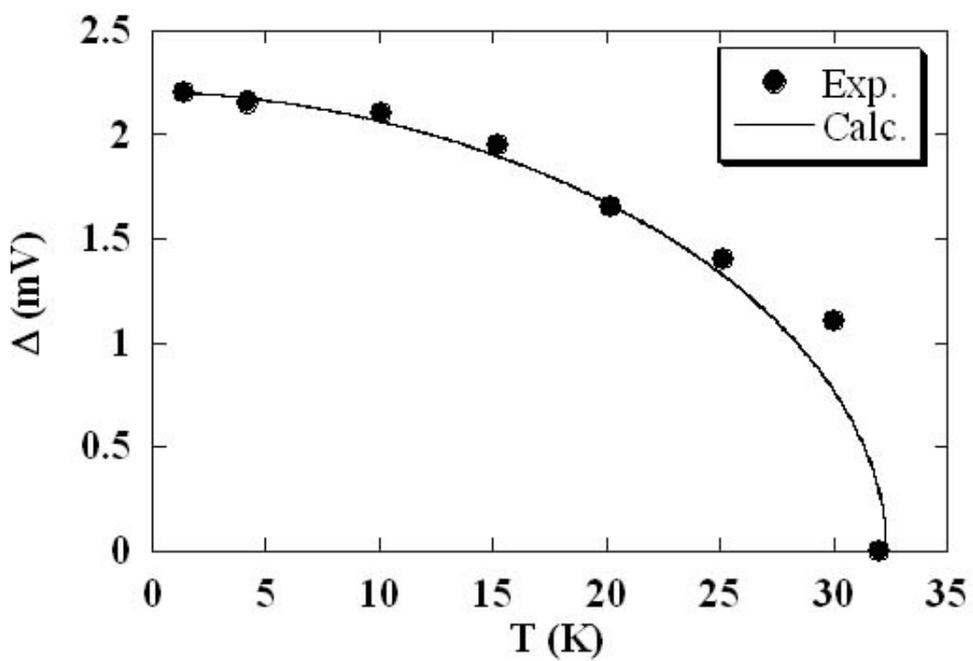

Fig. 5

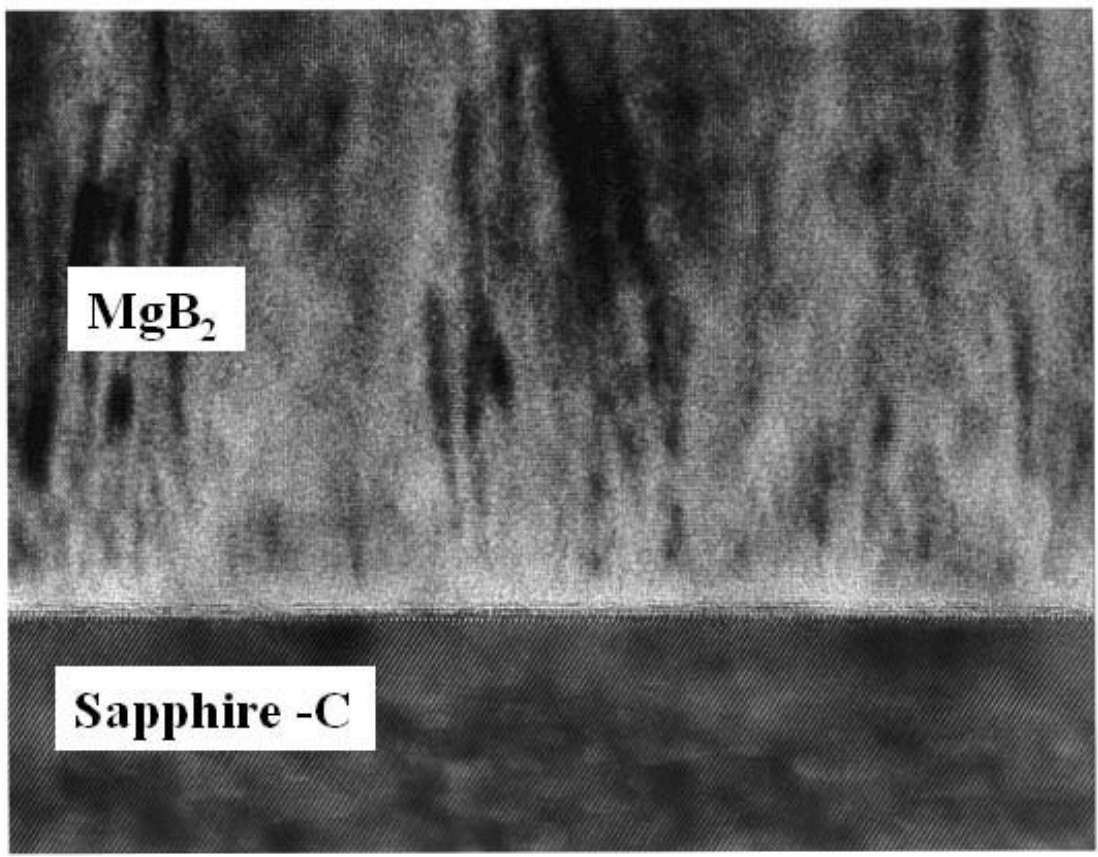

MgB₂

Sapphire -C

— 5 nm

Fig. 6

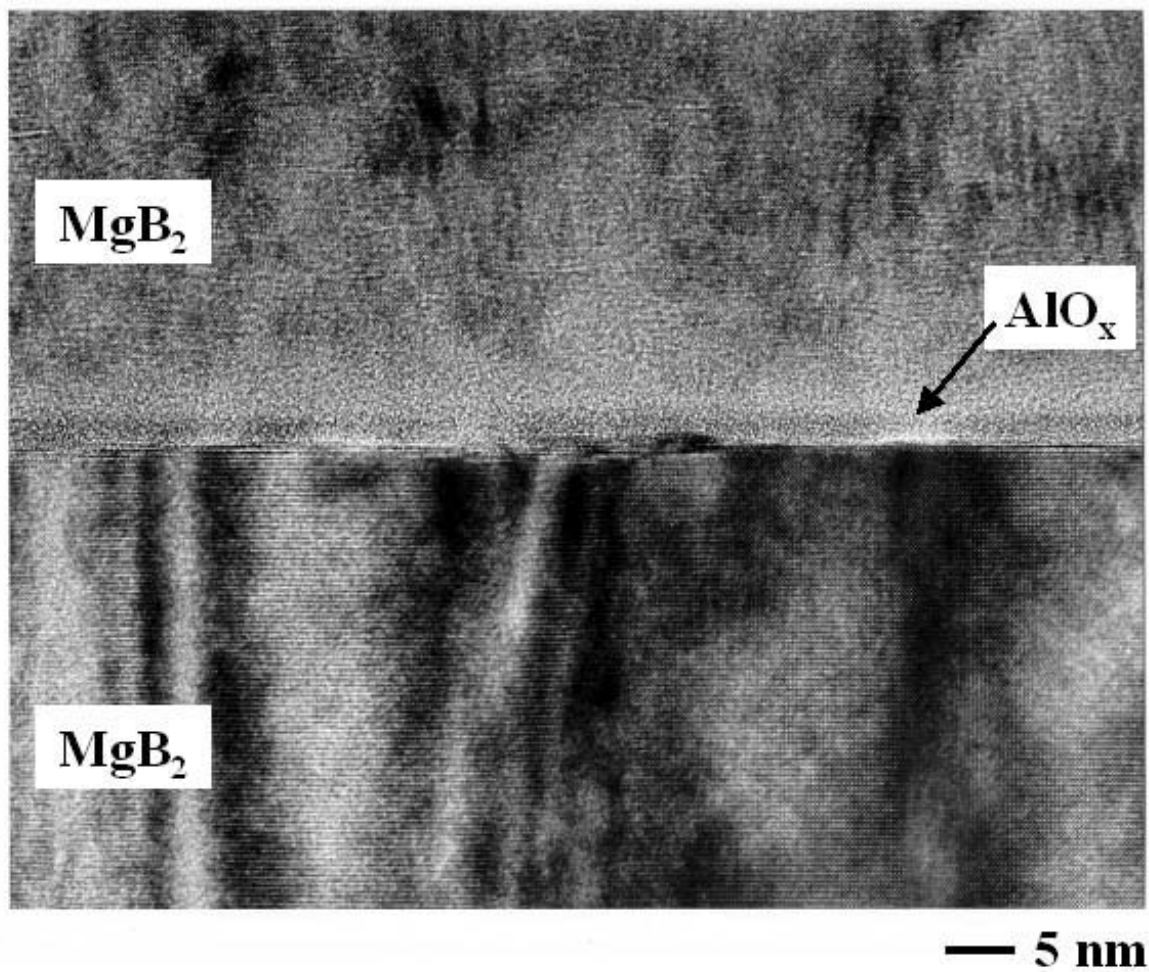

Fig. 7

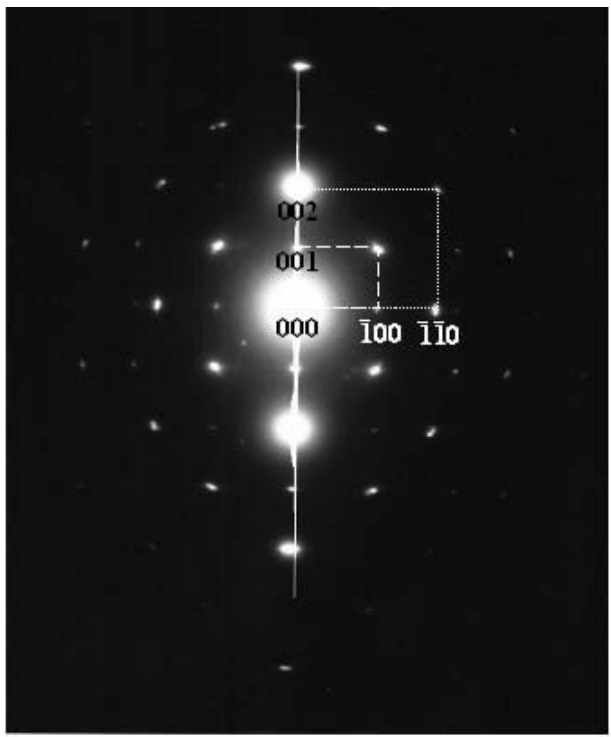

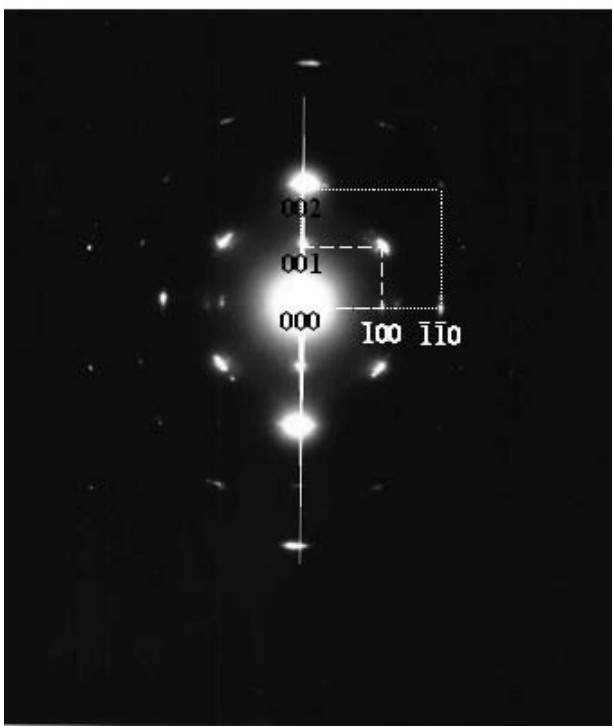

Fig. 8

| | $I_c$ ($\mu$A) | $R_N$ ($\Omega$) | $I_cR_N$ (mV) | $J_c$ (A/cm$^2$) | $R_NA$ ($\Omega$ cm$^2$) | $R_{sg}/R_N$ |
|---|---|---|---|---|---|---|
| A (100$\mu$m□) | 12.5 | 61 | 0.76 | 0.125 | $6.1*10^{-3}$ | 5.4 |
| B (25$\mu$m□) | 0.4 | 830 | 0.33 | 0.064 | $5.2*10^{-3}$ | 21.8 |
| C (25$\mu$m□) | 40.8 | 32 | 1.31 | 6.53 | $2.0*10^{-4}$ | 2.5 |

(at 4.2 K)

Table 1